# Key parameters generation of the navigation data of GPS Simulator

Tao Feng, *Qianqian Wu, Haipeng Zhang
School of Electronics & Information, Hangzhou Dianzi University,
Hangzhou, China, 310018
Email: qian3123534680@163.com

**Abstract.** The development of the GPS (Global Positioning System) signal simulator involving to a number of key technologies, in which the generation of navigation message has important significance. Based on analysis of the structure of GPS navigation data, the paper researches the production of telemetry word and handover word, parity check code, time parameters and star clock. Using disturbing force equation and Lagrange planetary motion equation extrapolate ephemeris parameters whose feasibility is verified through the Matlab software finally.

* Corresponding Author:
Qianqian Wu
ZhangSchool of Electronics & Information, Hangzhou Dianzi University, Hangzhou, China, 310018
Email: qian3123534680@163.com

## 1. Introduction

In order to test the static and dynamic performance of the GPS receiver, a GPS signal simulator to emulate real satellite signals in all sorts of environments is needed. The search had been started early in the foreign countries, and achieved great success [1]. But because it involves the sensitive technology, published literature very little, the research starts late in our country, especially the development of dynamic simulator [2]. In order to facilitate the simulator development, based on the ICD-GPS-200[3] (GPS interface control document) , this paper is organized as follows, section 2 summarizes the structure of the GPS navigation data, section 3 discusses the generation of some key parameters involved in the development of signal simulator and finally analyzes simulation results to verify the feasibility.

## 2. Navigation data structure

GPS navigation data is the basis of orientation and navigation whose speed is 50 b/s. Complete navigation data includes 25 pages (frames). Each page has 1500 bits, divided into 5 son pages (or called child frames).And every child frame contains 300 bits, 6 seconds of transmission time. The 1, 2, 3 child frame of each frame are the same, content is updated on per hour which launch satellite clock correction parameters and ephemeris, etc. The 4, 5 child frames have 25 forms (the same structure, different data), representing the satellite almanac, the ionosphere parameters etc [4]. 4, 5 sub-frame are 12.5 minutes finished first then repeat, content only update in the new navigation data. GPS navigation data structure is shown in **Figure1.** Transmission of a complete navigation data needing different 25 frames (12.5 min).

| TLM | HOW | satellite clock parameters |
|-----|-----|---------------------------|
| TLM | HOW | ephemeris parameters |
| TLM | HOW | satellite almanac |

Figure1. GPS navigation data structure

## 3. Key parameters generation

### 3.1 Telemetry word (TLM) and handover word (HOW)

Each child frame contains 10 words and each word has 30 bits. A child frame is always beginning with two special words, namely telemetry word (TLM) and handover word (HOW).

TLM is the first word in each child frame including synchronization head (10001011) in 8 bits, reserved 16 bits (simulator can be set 0) and parity. Synchronization head is used for testing the beginning of a child frame.

HOW includes GPS weeks information and 3 bits which represents a child frame ID.Z count (17bits) calculate a week time with 6 seconds for the unit (equal to a child frame duration),which plays a key role in the calculation of the satellite signal accurate launch time.

Z count is the starting time of the next child frame, counting from 0 count, every six seconds add 1, count to 100799 and then back to the 0 .because of a week from Sunday's 00:00:00 to Saturday's 23:59:59 cycle, being 604800 seconds. Z count multiplied by 6 minus 4.8 equals the starting time of next word in current frame. Production of Z count is as follows [1]:

$$Z = INT(t_{sec}/6) + 1 \quad (1)$$

$$t_{sec} = (d_1 \% 7) \times 86400 + h \times 3600 + m_m \times 60 + s \quad (2)$$

Where
$$d_1 = (y-1980) \times 365 + d_{oy}(m-1) + d + D_1 - 6 \quad (3)$$

$$D_1 = \frac{y-1980}{4} + 1 \quad (4)$$

$$d_{oy}(12) = \{0, 31, 59, 90, 120, 151, 181, 212, 243, 273, 304, 334\} \quad (5)$$

if $t_{sec} < 0$, $t_{sec} = t_{sec} + 86400$;
if $(y-1980)\%4 = 0$ or $m \leq 2$, $D_1 - 1$;

$y$, $m$, $d$, $h$, $m_m$, $s$ in the murderer represent the current year, month, day, minute, second and $t_{sec}$ is the number of seconds in a week with GPS time.

The 3 bits representing a child frame ID are accordingly set to 000,001,010,011,100,101 according to the number of current child frame (1-5).

### 3.2 Generation of parity inspection

GPS satellite transmits every "word" in which 25 to 30 bits representing parity inspection, they follow error detection criteria of Hamming code (32, 26). Because of GPS navigation data with 24 parity bits, (32, 26) Hamming code can remove two bits to constitute (32, 24) shorten code, the correction ability and minimum distance are the same with original code. In this case, shorten Hamming code has the check matrix H:

$$H = \begin{pmatrix} 1 & 1 & 1 & 0 & 1 & 1 & 0 & 0 & 0 & 1 & 1 & 1 & 1 & 0 & 0 & 1 & 1 & 0 & 1 & 0 & 0 & 1 & 0 \\ 0 & 1 & 1 & 1 & 0 & 1 & 1 & 0 & 0 & 0 & 1 & 1 & 1 & 1 & 0 & 0 & 1 & 1 & 0 & 1 & 0 & 0 & 1 \\ 1 & 0 & 1 & 1 & 1 & 0 & 1 & 1 & 0 & 0 & 0 & 1 & 1 & 1 & 1 & 0 & 0 & 1 & 1 & 0 & 1 & 0 & 0 \\ 0 & 1 & 0 & 1 & 1 & 1 & 0 & 1 & 1 & 0 & 0 & 0 & 1 & 1 & 1 & 1 & 0 & 0 & 1 & 1 & 0 & 1 & 0 \\ 1 & 0 & 1 & 0 & 1 & 1 & 1 & 0 & 1 & 1 & 0 & 0 & 0 & 1 & 1 & 1 & 1 & 0 & 0 & 1 & 1 & 0 & 1 \\ 0 & 0 & 1 & 0 & 1 & 1 & 0 & 1 & 1 & 1 & 0 & 1 & 0 & 1 & 0 & 0 & 0 & 1 & 0 & 0 & 1 & 1 & 1 \end{pmatrix}$$

The result of calibration is the matrix $S(6 \times 1) = H(6 \times 24) \oplus M(1 \times 24)$ ($\oplus$ is xor operation) .That is, the 1 ~ 24 bits in every word respectively has xor operation with each line in H. Each line get 1 bit representing calibration result, so the final calibration results are 25 to 30 bits.

### 3.3 The time parameters

There are four significant time parameters: toc, toe, $I_{ODC}$, $I_{ODE}$ located in subframes 1,2,3.

$t_{oc}$, $t_{oe}$ represent the first data block and ephemeris reference time since each week starting epoch(Saturday / Sunday night zero). In a simulator, both values are the same, which is generated as follows:

$$t_{oc} = t_{oe} = (d1\%7) \times 86400 + (h - Timezone) \times 3600 + m \times 60 + s + 13 \quad (6)$$

$$t_{oc}' = \frac{t_{oc}}{16}, t_{oe}' = \frac{t_{oe}}{16} \quad (7)$$

The parameter value are the same with formula (3) - (5), $t_{oc}'$,$t_{oe}'$ are ephemeris value by $t_{oc}$, $t_{oe}$ coefficient conversion.

Satellite clock data age IODC is the extrapolation time intervals of clock correction. As time goes on, the satellite clock correction parameters precision will be decreased, so the clock data age is mainly used for confidence evaluation of clock correction parameters。It is generated as follows:

$$IODC = t_{oc} - t_L \quad (8)$$

Where $t_L$ is the final observation time to calculate the clock correction parameters.

IODE is the extrapolation time interval of satellite ephemeris. The smaller value, the higher confidence. It is generated as follows:

$$IODE = t_{oe} - t_l \quad (9)$$

Where $t_l$ is final observation time used to calculate the broadcast ephemeris measure.

### 3.4 Star clock parameters

The child frame 1 contains satellite clock information, which is used to determine when the satellite navigation launches. The master atomic clock of the ground is reference of GPS system. A GPS satellite clock and GPS time system has difference, needing to be corrected. Satellite clock correction is as follows:

$$\Delta t_a = a_0 + a_1(t - t_{oc}) + a_2(t - t_{oc})^2 \quad (10)$$

Where $t_{oc}$ is reference moment of first data block, position in 9-24 bits of the eighth word; t is the calculation moment of the satellite clock error; $a_0$ is time offset(second) of satellite clock errors, 1-8 bits of the ninth word; $a_1$ is satellite clock rate coefficient of deviation(SEC/sec),9-25bit bits of the ninth word; $a_2$ is drift coefficient of satellite clock speed rate(SEC/sec$^2$),1-22 bits of tenth word.

Because of each channel signal of GPS simulator is produced by the same frequency clock, so the clock parameters of each satellite in simulator are the same. By comparing with high-precision atomic clock regularly get the newest clock frequency deviation and frequency drift as the initial value of clock correction parameters extrapolation in $t_{oc}$, which instead of the former clock parameters[5]. After predetermined warm-up time when frequency clock stability to the requirement, use the local time $t_0$ as the systemic starting reference time (the corresponding GPS time is $t_{GPS0}$).We can set the value of $a_0$, $a_1$, $a_2$ at the moment $t_{GPS0}$:

$$a_0(t_{GPS0}) = 0 \quad (11)$$

$$a_1(t_{GPS0}) = a_1 \quad (12)$$

$$a_2(t_{GPS0}) = a_2 \quad (13)$$

So,

$$a_0(t_{oc}) = a_1(t_{oc} - t_{GPS0}) + a_2(t_{oc} - t_{GPS0})^2 \quad (14)$$

$$a_1(t_{oc}) = a_1(t_{GPS0}) + a_2(t_{oc} - t_{GPS0}) \quad (15)$$

$$a_2(t_{oc}) = a_2 \quad (16)$$

The clock correction parameters at $t_{oc}$ moment can be converted binary value and wrote in navigation.

### 3.5 Ephemeris parameters extrapolation

GPS satellite ephemeris a set of orbital elements and the rate of change at a certain moment. Satellite position and velocity at any one time can be calculated according to the ephemeris.Sub frame 2 and 3 release of GPS broadcast ephemeris, which include Kepler orbit elements and the necessary orbit perturbation correction parameters relative to some reference epoch. Correcting known reference ephemeris by orbital elements of perturbation then any observation epoch of satellite ephemeris can be extrapolated. There are 16 GPS broadcast ephemeris parameters, including ephemeris reference epoch, Kepler six parameters such as a,$e_s$,$i_0$,$\Omega_0$,$\omega$,M and nine perturbation parameters $\Delta n$, $\dot{\Omega}$, $\dot{I}$, Cuc, Cus, Crc, Crs, Cic, Cis. They are defined as follows [6]:

$t_{oe}$– ephemeris reference epoch;

a–the long axle of the elliptical orbit;
$e_s$–the elliptical orbit eccentricity;
i0–the orbital inclination angle;
$\Omega_0$–right ascension of the ascending node;
ω– angular distance of perigee;
M–mean anomaly;
Δn–average angular speed difference between precise ephemeris and given parameters;
$\dot{\Omega}$ –right ascension of the ascending node change rate;
I–the orbital inclination change rate;
Cuc, Cus–cosine, sine harmonic correction term amplitude of latitude amplitude;
Crc, Crs–cosine, sine harmonic correction term amplitude of orbit radius;
Cic, Cis–correction of the amplitude of sinusoidal, cosine harmonic

Generation of GPS broadcast ephemeris is mostly fitting method [7], use precise ephemeris given satellite position as observations fitting the broadcast ephemeris parameters. The fitting accuracy of this method is very high, between 2h to 3h but with the fitting time increases, precision obviously decreased.

Satellite simulator ephemeris is so long as keep the relative position of the satellite in each orbital plane meeting the requirements [8], without requiring satellite position the same as its real value at any time. We therefore adopt another method to generate the broadcast ephemeris. Consider the perturbed motion of the satellite, the satellite orbital elements change over time. In all perturbation, the influence of the gravitational field perturbation forces is the largest. Higher due to the GPS satellite orbits, generally only consider the perturbation caused by the second-order spherical harmonic coefficients, to meet the orbit accuracy requirements. So the perturbation function is expressed as:

$$\Delta V = -\frac{3}{2}\frac{\mu}{\gamma}J_2(\frac{a_e}{\gamma})^2 \times \left[\left(\frac{1}{3}-\frac{1}{2}\sin^2 i\right)+\frac{1}{2}\sin^2 i \cos 2(\omega+f)\right] \quad (17)$$

Where $\Delta V$ represents satellite gravitational potential, *f* is the true anomaly angle and *r* is distance between the satellite and the center of the earth. Expand $\Delta V$ on the orbital parameters according to the Lagrange planetary equations derivation and the six orbital parameters at time t can be got as follows [9]:

$$a(t) = a(t_0) + \frac{\partial a}{\partial t}(t-t_{oc}) = a(t_0) \quad (18)$$

$$e(t) = e(t_0) + \frac{\partial e}{\partial t}(t-t_{oc}) = e(t_0) \quad (19)$$

$$i(t) = i(t_0) + \frac{\partial i}{\partial t}(t-t_{oc}) = i(t_0) \quad (20)$$

$$\Omega(t) = \Omega(t_0) + \frac{\partial \Omega}{\partial t}(t-t_{oc}) = \Omega(t_0) - \frac{3}{2}a_e^2\frac{J_2}{p^2}(n+\Delta n)\cos i(t-t_0) \quad (21)$$

$$\omega(t) = \omega(t_0) + \frac{\partial \omega}{\partial t}(t-t_0) = \omega(t_0) - \frac{3}{2}a_e^2\frac{J_2}{p^2}(n+\Delta n)\left(2-\frac{5}{2}\sin^2 i\right)(t-t_0) \quad (22)$$

$$M(t) = M(t_0) + \frac{\partial M}{\partial t}(t-t_0) = M(t_0) - \frac{3}{2}a_e^2\frac{J_2}{p^2}(n+\Delta n)\left(-1+\frac{3}{2}\sin^2 t\right)(1-e^2)(t-t_0) \quad (23)$$

Where,
$$p = a(1-e^2) \quad (24)$$
$$a_e = 6378137m \quad (25)$$
$$J_2 = 108263 \times 10^{-8} \quad (26)$$

$$\mu = 3986005 \times 10^8 \left( m^3 / s^2 \right) \quad (27)$$

$$n_0 = \sqrt{\frac{\mu}{a}} \quad (28)$$

$R$ is earth's equatorial radius.

So the parameters extrapolation of the follow-up time can be got on the basis of a certain amount of known ephemeris. A GPS receiver working continuously for 24h, the complete collection of the healthy GPS navigation data can be got, then demodulate the data to get a group of ephemeris which is the basis of ephemeris calculation of any time.

## 4. Simulation
### 4.1 Reading GPS ephemeris data based on Matlab

In order to be able to use data from different receiver, currently adopted a receiver independent receiver independent exchange format RINEX (Receiver Independent Exchange Format), a pure ASCII code text file. Different time ephemeris parameters can be obtained according to the GPS navigation message files in RINEX format. Based on Matlab software build structure array navdata, a total of 38 fields whose type are NaN. Store data to the corresponding field name with easy access to satellite parameters when reading RINEX navigation file [10]. The structure array navdata can be built as follows:

navdata=struct('prn',NaN,'year',NaN,'month',NaN,'day',NaN,'hour',NaN,'minute',NaN,'second',NaN,'a0',NaN,'a1',NaN,'a2',NaN,'iode',NaN...'crs',NaN,'deltan',NaN,'M0',NaN,'cuc',NaN,'e',NaN,'cus',NaN...'toe',NaN,'cic',NaN,'comega',NaN,'cis',NaN,'i0',NaN,...'omega',NaN,'comegadot',NaN,'idot',NaN,'iodc',NaN)

Read the navigation file, through the loop line to find the special string "END OF HEADER", navigation file header files and ephemeris data dividing point, marking the start position of acquiring ephemeris. Then get ephemeris data and store to the array structure. Part of the specific code as follows:

```
fid=fopen('E:\ephemerisfile.txt','r');
head_lines=0;
while 1
head_lines=head_lines+1;
line=fgetl(fid);
answer=findstr(line,'END OF HEADER');
if~isempty(answer)
break
end
end
noeph=-1;
while 1
noeph=noeph+1;
line=fgetl(fid);
if line==-1
break
end
end
noeph=noeph/8;
frewind(fid);
for l=1:head_lines
line=fgetl(fid);
end
for i=1:noeph
line=fgetl(fid);%first line
y=str2num(line(3:6));
if y>79
navdata(i).year=y+1900;
else
navdata(i).year=y+2000;
end
navdata(i).month=str2num(line(7:9));
```

```
navdata(i).day=str2num(line(10:12));
navdata(i).hour=str2num(line(13:15));
navdata(i).minute=str2num(line(16:18));
navdata(i).second=str2num(line(19:22));
navdata(i).af0=str2num(line(23:41));
navdata(i).af1=str2num(line(42:60));
navdata(i).af2=str2num(line(61:79));
line=fgetl(fid);%second line
navdata(i).iode=str2num(line(4:22));
navdata(i).crs=str2num(line(23:41));
navdata(i).deltan=str2num(line(42:60));
navdata(i).M0=str2num(line(61:79));
line=fgetl(fid);%third line
navdata(i).cuc=str2num(line(4:22));
navdata(i).e=str2num(line(23:41));
navdata(i).cus=str2num(line(42:60));
navdata(i).sqrta=str2num(line(61:79));
…
```

### 4.2 The simulation results and analysis

Collect the GPS satellites navigation message by a GPS receiver working continuously for 24h. Read the RENIX file to get the star ephemeris parameters through Matlab. Six orbit parameters can be calculated on basis of the parameters gotten. Taking no. 1 GPS satellite as an example, as shown in Table 1, orbit parameters generating method of other satellite are the same. t1 is 2011-06-28T 0:00:00 and t2 for 2011-06-30T 4:00:00 in **Table 1**, the parameters value of the t2 time can be generated by parameters extrapolation of t1.

**Table 1.** M, $\Omega$, $\omega$ extrapolation

|  | mean anomaly M | right ascension of the ascending node $\Omega$ | argument of perigee $\omega$ |
|---|---|---|---|
| t1(0：00：00) | -0.168422434376D+01 | 0.154255529548D+01 | 0.148394519733D+00 |
| t2(4：00：00) | 0.716100835560D+00 | 0.259243759787D+01 | 0.149925811275D+01 |
| t2 extrapolation | 0.71608555028638 | 2.59244182259716 | 1.4992494803166 |

Taking the value of $\Omega$ as an example, the contrast between the received value and the parameters extrapolation at time 2011-06-30T 0:00:00 to 2011-06-30T 14:00:00 moment are shown in **Figure 2**. The blue marker represents the received value and pink curve for 01 satellite in celestial the value of parameters extrapolation.

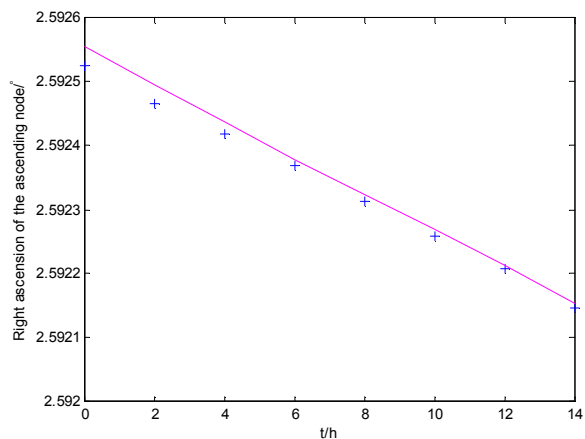

**Figure 2.** contrast between the received value and the extrapolation of $\Omega$

In order to verify the validity of the above-mentioned parameters extrapolation, in Mat lab simulation .Using the above extrapolation of parameters at 2011-06-30T 8:00:00, simulation of the GPS constellation maps and satellite orbit variation through the Mat lab software are Shown in **Figure 3, 4** [11].Figure 3 shows that the distribution of the 24 satellites in six orbital planes, each orbital planes separated by 60 °, namely the ascending node right ascension difference of each orbit is 60 °. The satellite orbital inclination is 55 °. So satellite constellation distribution and operating rules comply with the real situation. Pink curve for 01 satellite in celestial coordinates position, blue curve for in orbit right angle coordinate system position are shown in figure 4.

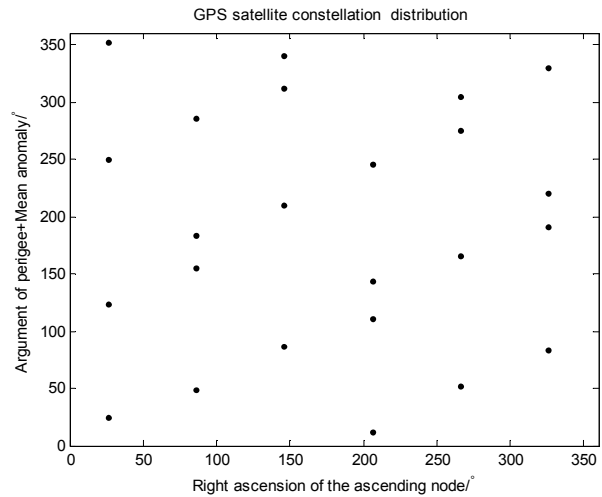

**Figure 3.**GPS constellation maps

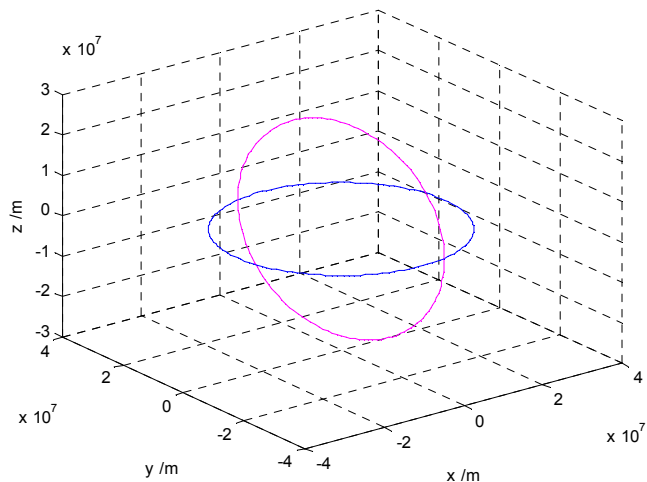

**Figure 4.**Satellite orbit variation

3D error amplification factor PDOP often used to measure the location of the user's measurement accuracy [12]. The value of PDOP obtained by extrapolation of ephemeris parameters are shown in **Figure 5**, which ranging between 1.5 and 3.5. In the simulation process, calculate a set of orbital parameters every 300 seconds. So the extrapolation of the orbital parameters can be used for the receiver.

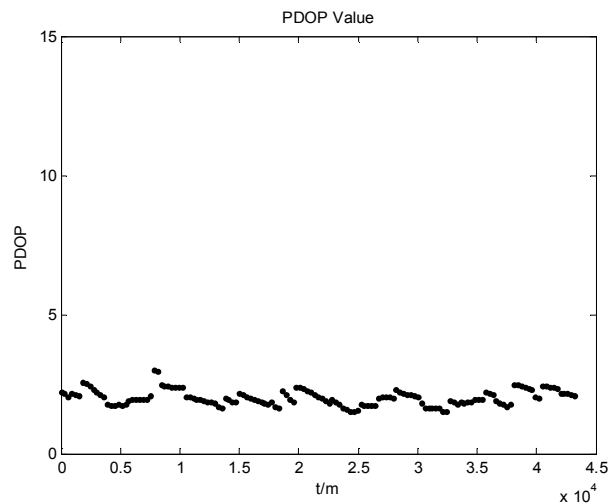

Figure 5.value of PDOP

## 5. Conclusions

   The generation of navigation data has important significance in development of GPS simulator. This paper studies some key parameters, such as parity check code, clock parameters, ephemeris parameters extrapolation of GPS navigation data and relevant generation algorithm.Using perturbation equation and Lagrange planetary motion equation deduced ephemeris parameters,which meeting the performance requirements of GPS signal simulator.
   In this paper the ephemeris extrapolation only consider the perturbation caused by the second-order spherical harmonic coefficients. In the future we can consider higher order perturbation influence on the satellite ephemeris parameters extrapolation.

## 6. Acknowledgement

   The authors are grateful to the support of national nature science funds (60902040).

## References
[1]   Alison Brown, Neil Gerein, "Advanced GPS hybrid simulator architecture", Proceedings of ION 57 Annual Meeting, pp.1-8, 2001.
[2]   Liu-LI Li, Wang-Ke Dong, "The Trend of Satellite Signal Simulators", Engineering of surveying and mapping, Gnss World of China,China Research Institute of Radio Wave Propagation,vol.17,no. 3, pp. 58–62 ,2010.
[3]   GPS-ICD-200C Interface Control Document [R], ARINC Research Corporation, 1993.
[4]   Zhang-Wen and M.Ghogho, "Hypothesis testing analysis and unknown parameter estimation of GPS signal detection", Journal of Central South University,vol.17, no. 5, pp. 1290–1301, 2012.
[5]   Wu-Jing,Chang-Qing,Wu-Jing Pei and Zhang-Qi Shan,"Generation of High dynamic GPS signal simulator satellite star calendar",Radio Engineering of China,The fifty-fourth Research Institute of China Electronic Technology Group Corporation.,vol.34, no. 5, pp. 42–43+60, 2004.
[6]   Zhang-Yu Xiang, Satellite orbit Determination, National defence industry press, Beijing, 2006.
[7]   Cui-Xian Qiang, Jiao-Wen Hai and Qin-Xian Ping, "Discussion on the fitting of GPS broadcast ephemeris parameters", Science of Surveying and Mapping, China surveying and mapping institute of science, vol.31, no. 1, pp.25-26+48+3, 2006.
[8]   Rajan JA, OrrM, "On-orbit Valid action of GPS-Autonomous Navigation[C]", Proceedings of the ION59th Annual Meeting, Albuquerque, pp.411-419, 2003.
[9]   Zhang-Bo Chuan, Chang-Qing, Zhang-Qi Shan and Kou-Yan Hong. "Research on the production of the navigation data of high dynamic GPS signal simulator ",Journal of Beijing University of Aeronautics and Astronautics, Beijing university of aeronautics and astronautics, vol.3,no. 3, pp. 284–287,2005.
[10]  Zhang-Ni and Wang-Biao Biao. "Reading standard RINEX format for GPS ephemeris data based on Matlab", Electronic Design Engineering, Xi'an City Sancai Technology Industrial Company Limited vol.18, no. 8, pp.23-25.2010.

[11]   Du-XinXin and Zhang-Qi Shan. "Algorithm study of satellite motion status for GPS signal simulator", Electronic Measurement Technology, Beijing radio technology research institute.vol.30,no. 7, pp. 112–114 ,2007.

[12]   Dong-Fang,Jia-Chuan Ying and Gu-Wei,"A Study of the Reltion Between GDOP and Gu", Journal of Dalian Marine College, Dalian Maritime University,vol.19,no. 2, pp. 210–218 ,1993.

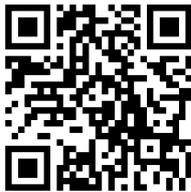
Free download this article and more information